\documentclass{aa}
\usepackage{txfonts}
\usepackage{graphicx}
\begin{document}

\title{A binary star sequence in the outskirts of the disrupting Galactic open cluster UBC\,274}

\author{Andr\'es E. Piatti\inst{1,2}\thanks{\email{andres.piatti@unc.edu.ar}}}

\institute{Instituto Interdisciplinario de Ciencias B\'asicas (ICB), CONICET UNCUYO, Padre J. Contreras 1300, M5502JMA, Mendoza, Argentina;
\and Consejo Nacional de Investigaciones Cient\'{\i}ficas y T\'ecnicas (CONICET), Godoy Cruz 2290, C1425FQB,  Buenos Aires, Argentina
}

\date{Received / Accepted}

\abstract{
We report the identification of a numerous binary star population in the recently
discovered $\sim$ 3 Gyr old open cluster UBC\,274. It becomes visible once
the cluster color-magnitude diagram is corrected by differential reddening
and spans mass ratios ($q$) values from 0.5 up to 1.0. Its stellar density
radial profile and cumulative distribution as a function of the distance from
the cluster's center reveal that it extends out to the observed boundaries
of the cluster's tidal tails ($\sim$ 6 times the cluster's radius) following a
spatial distribution indistinguishable from that of cluster Main Sequence (MS)
stars. Furthermore, binary stars with $q$ values smaller or
larger than 0.75 do not show any spatial distribution difference either. 
From {\it Gaia} DR2 astrometric and kinematics data we computed
Galactic coordinates and space velocities with respect to the cluster's center
and mean cluster space velocity, respectively. We found that, cluster members located all along the tidal tails, irrespective of  being a single or binary star, 
move relatively fast. The projection of their motions on the  Galactic plane resembles that of a rotating solid body, while those along the radial direction 
from  the Galactic center and perpendicular to the Galactic plane suggest that 
the cluster is being disrupted. The similarity of the spatial distributions and kinematic patterns of cluster MS and binary
stars reveals that UBC\,274 is facing an intense process of
disruption that has apparently swept out any signature of internal dynamic
evolution like mass segregation driven by two-body relaxation.
}
 

\titlerunning{The binary star sequence of UBC\,274}

\authorrunning{Andr\'es E. Piatti}

\maketitle

\markboth{Andr\' es E. Piatti: }{The binary star sequence of UBC\,274}

\section{Introduction}

UBC\,274 (R.A.: 10$^h$24$^m$46.50$^s$, Dec.: -72$^o$34'28.36", {\it l:} 292$^o$. 3622, b: -12$^o$.7920) is a Galactic open cluster recently discovered by 
\cite{castroginardetal2020}
from astrometric, kinematics and photometric data sets available at the {\it Gaia} DR2  archive\footnote{http://gea.esac.esa.int/archive/} \citep{gaiaetal2016,gaiaetal2018a,gaiaetal2018b}. They recognized the new object
after applying a machine learning based technique and a deep learning
artificial neural network. It resulted to be a $\sim$ 3 Gyr old open cluster 
 (see the cluster color-magnitude diagram (CMD) built by them in 
Fig,~\ref{fig:fig1}) located at  
$\sim$ 2 kpc from the Sun. From 365
{\it bonafide} members, they found that the cluster has a remarkable 
elongated shape, from which they concluded that it is being the subject of an
extensive tidal disruption process. 

Because of its populous extra-tidal features, the cluster deserves much more of our 
attention. For instance, an analysis of the spatial distribution of the different 
stellar populations in the cluster CMD  would contribute to 
know its ongoing internal dynamical evolutionary stage
\citep[see, e.g.,][]{angeloetal2019b,piattietal2019c}. The relatively deed and complete cluster CMD
could also be useful to dive into the still debatable existence of
extended Main Sequence (MS) turnoffs in Galactic open clusters and  their
origin \citep[see, e.g.,][]{cordonietal2018,pb2019,dejuanovelaretal2020}. The position and shape of the
observed tidal tails could be used to constrain models of the formation of substructures
along them \citep[see, e.g.,][]{montuorietal2007,kupperetal2010}, among others.

In this work, we closely revisited the {\it Gaia} DR2 data set for the 365 cluster
members identified by  \citet{castroginardetal2020} and found that UBC\,274
contains a  well populated binary sequence, that extends out to the
observed outskirts of the disrupting cluster following similar spatial and
kinematical distributions as escaping single cluster MS stars. If UBC\,274
internal dynamics  was driven only by two-body relaxation, its binary population 
should be more centrally concentrated than  that of the single stars. As far as we are aware, 
we report the first open cluster where different spatial distributions of single and
binary stars are not observed. The analysis is organized as
follows: in Section 2 we  present an intrinsic cluster CMD from which we identify the 
cluster binary star population, while in Section 3 we discuss its resulting
spatial and kinematical distributions. Finally, in Section 4 we summarize the
main outcomes of this work.

\section{The cluster binary star sequence}

\begin{figure}
\includegraphics[width=\columnwidth]{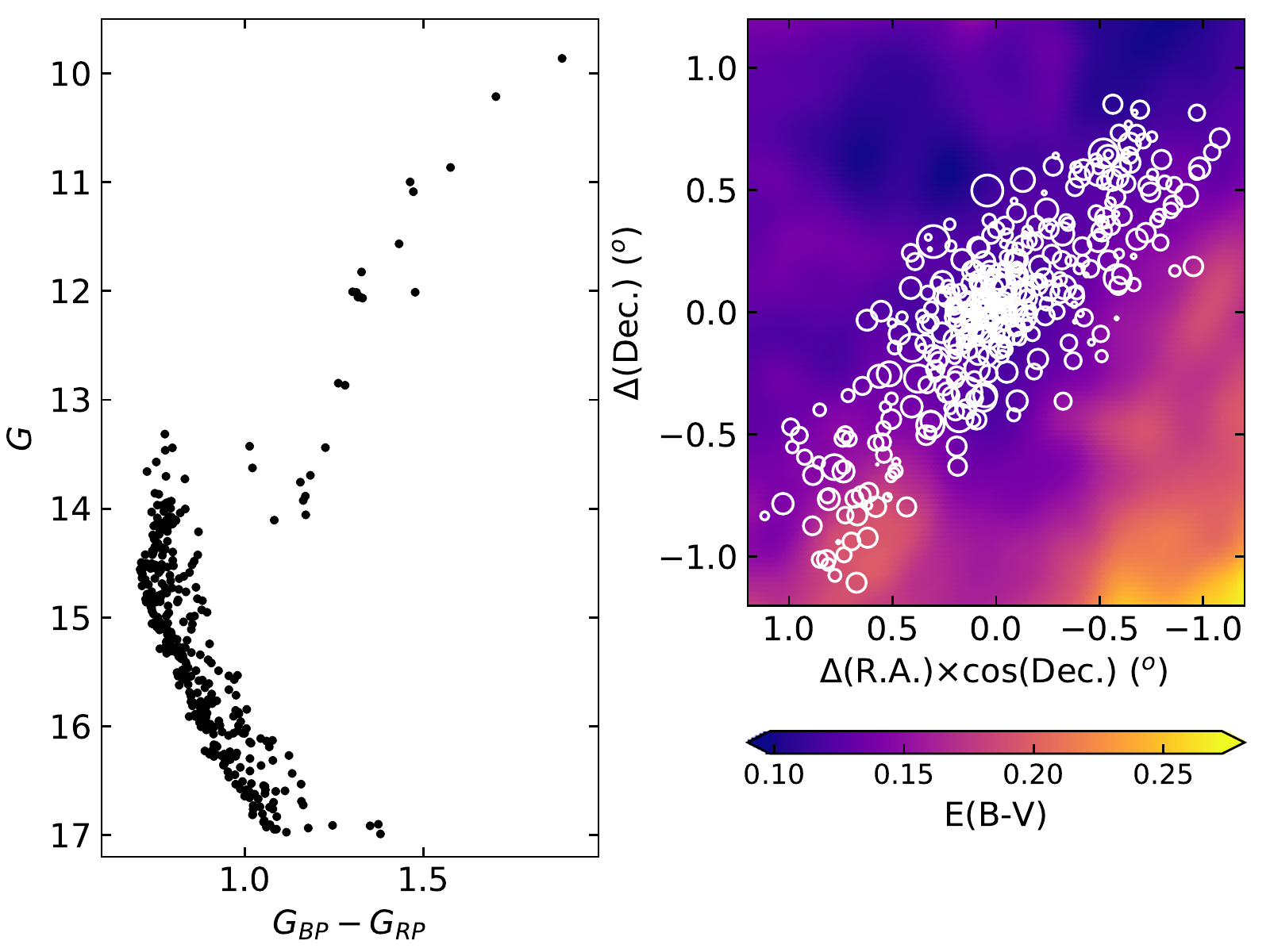}
\caption{ Observed cluster CMD (left panel).
Interstellar reddening ($E(B-V)$) map for the cluster field  (right panel). The
cluster members are represented by open circles whose sizes are
proportional to their $G$ brightnesses.}
\label{fig:fig1}
\end{figure}

\begin{figure}
\includegraphics[width=\columnwidth]{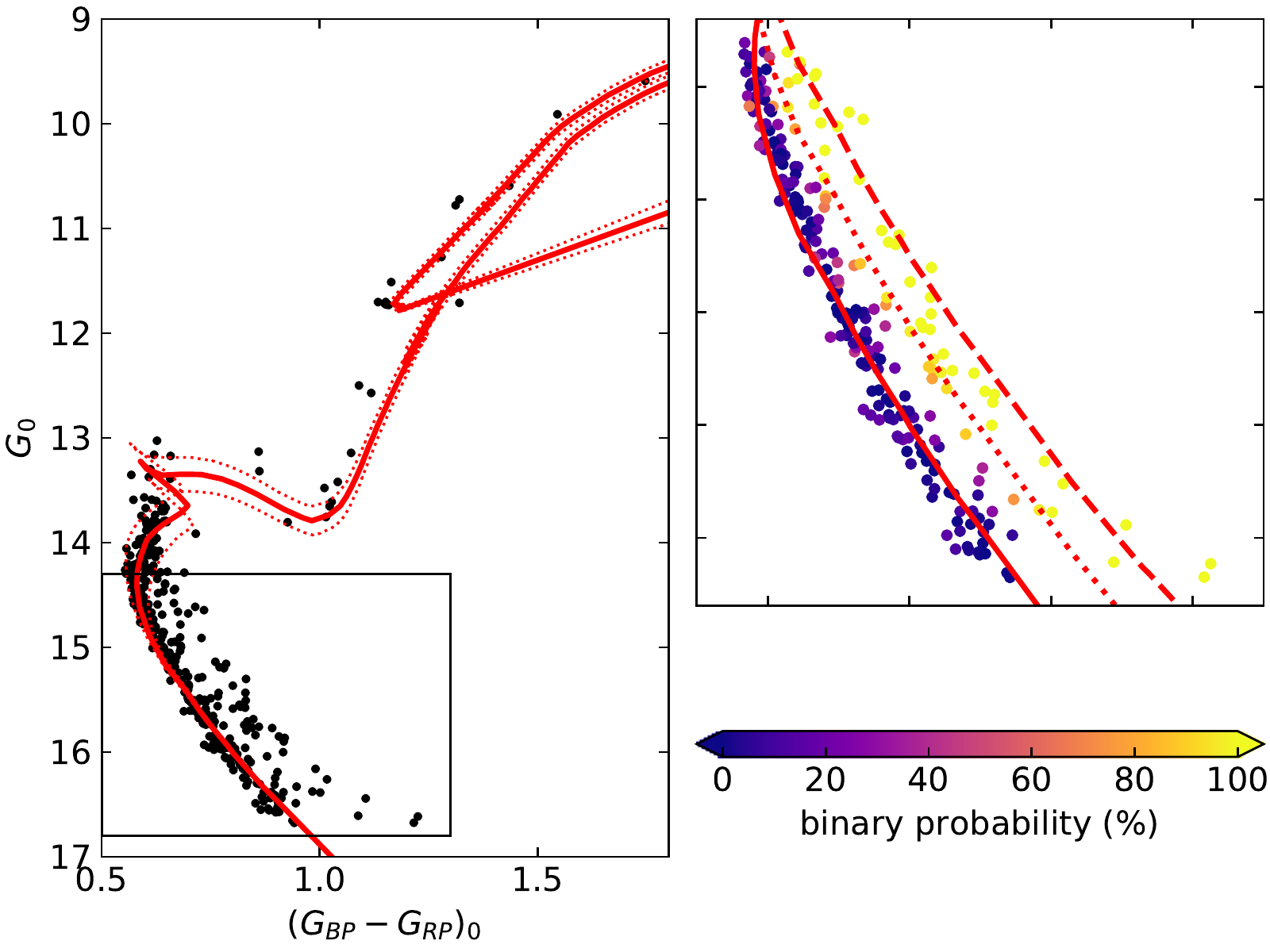}
\caption{Intrinsic cluster CMD (left panel). Isochrones of log($t$ /yr) =
 9.45 (solid line), 9.40 and 9.50 (dotted lines) are superimposed. The rectangle
 illustrates the zoom-in region depicted in the right panel, where the
 isochrone of log($t$ /yr) = 9.45 is superimposed for $q$= 0, 0.5 and 1.0 with solid, dotted and dashed lines,
 respectively.}
\label{fig:fig2}
\end{figure}

The presence of differential reddening across the field of star clusters can blur 
fundamental features of their CMDs. For this reason, we
first examined the spatial variation of the interstellar reddening across
UBC\,274. We built the interstellar reddening map by retrieving the 
 $E(B-V)$ values obtained by \citet{sf11} and provided by NASA/IPAC Infrared Science  Archive\footnote{https://irsa.ipac.caltech.edu/}.  Figure~\ref{fig:fig1} depicts the resulting spatial distribution of $E(B-V)$ values,
where we superimposed the loci of cluster members, represented
by open circles with sizes proportional to their $G$ brightnesses.
As can be seen, the cluster field is affected by some amount of
differential reddening. 
We then corrected by interstellar reddening the $G$ magnitude and 
$G_{BP} - G_{RP}$ color of each star using the   corresponding
individual $E(B-V)$ value
 according to the star's coordinates  in the reddening map and the relationships 
$A_G$ = 2.44 $E(B-V)$ and $E(G_{BP} - G_{RP})$= 1.27 $E(B-V)$
\citep{cetal89,wch2019}.  Figure~\ref{fig:fig2} shows the reddened corrected 
(intrinsic) CMD.

The observed cluster CMD clearly shows a red giant branch, a
red clump, a sub-giant branch, a nearly 4 mag long MS
and a binary star sequence spanning the whole MS magnitude
range.  After the reddening correction the binary star sequence is much more pronounced.
Hence, we emphasize in the importance of correcting the {\it Gaia} DR2 
photometry by differential reddening as a crucial step to make visible the 
well delineated and populous cluster binary star strip. Then, we
adopted the mean cluster parallax and the cluster metallicity from \citet{castroginardetal2020} and found that the  \citet{betal12}
theoretical isochrone  which best resembles the cluster
 features, i.e., position and shape of the different giant phases,
the MS turnoff and the curvature of the MS, is that of log($t$ /yr) = 9.45 
$\pm$ 0.05 (2.8$\pm$0.2 Gyr) (see Fig.~\ref{fig:fig2}, left panel). 

We statistically distinguished the cluster binary star population from the
cluster MS by running extensive Monte Carlo experiments. We
considered the photometric errors in $G$ and $G_{BP} - G_{RP}$
derived by \citet{evansetal2018} and the errors in $E(B-V)$ from 
\citet{sf11}, and used the isochrone of log($t$ /yr) = 9.45 as a
ridge line for the cluster MS from  $G_0$ = 14.3 mag down to 16.8
mag (see Fig.~\ref{fig:fig2}, right panel). In order to estimate the
probability for a star to belong to the cluster MS, we measured the
distance from that star to the cluster ridge line, adopting for
both end points (the star and the closest  position on the
cluster ridge line) the corresponding errors in  $G_0$ and $(G_{BP} - G_{RP})_0$.
We performed a thousand measurements of such a distance, allowing 
random values of magnitudes and colors within 3$\sigma$ for the
star and the position on the ridge line, respectively. Then, we
considered that a star belongs to the cluster MS 
 if accomplish the criterion: the distance to the MS is less than the 
sum of the errors along the line connecting the star and the ridge line. We finally obtained the probability ($P$) of a 
star to belong to the cluster MS dividing by 10 the number of times
it satisfied the above criterion. The difference  100 - $P$ gives the probability
of a star to be a binary star. This is because, all the stars are assumed
to be cluster member according to the membership criteria applied
by \citet{castroginardetal2020}.  Figure~\ref{fig:fig2} (right panel) shows 
color-coded binary probabilities. In the subsequent analysis we
considered binary stars those with $P$ < 40 per cent, which
roughly corresponds to secondary to primary mass ratios, $M_2/M_1 = q$
$\ga$ 0.5.

\section{Analysis and discussion}

\begin{figure}
\includegraphics[width=\columnwidth]{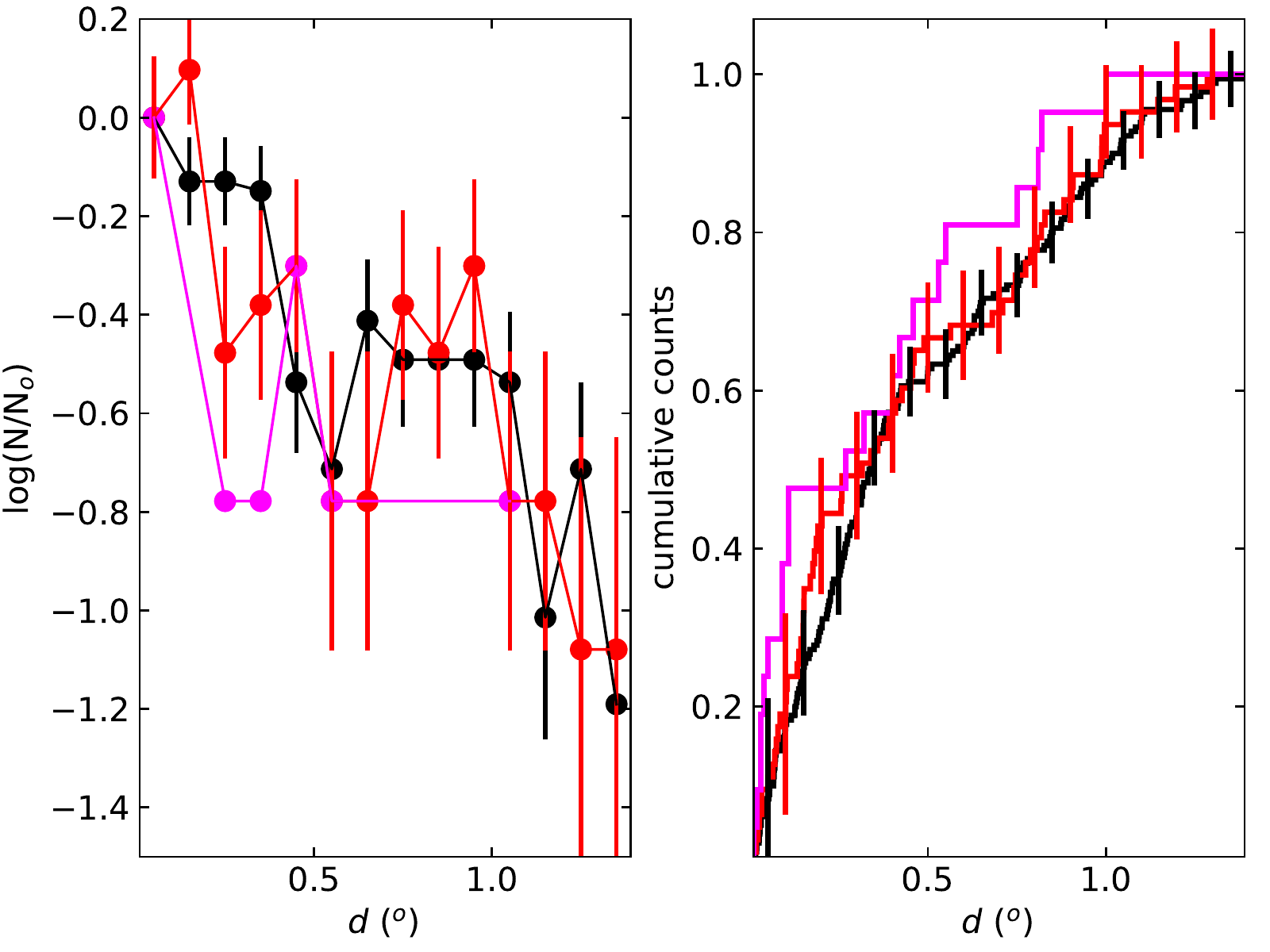}
\caption{Observed radial density profiles (left panel) and cumulative
distribution functions (right panel) for cluster MS (black lines),
binary (red lines),  and red giant stars  (magenta lines). Error bars are also indicated.}
\label{fig:fig3}
\end{figure}

Using the positions of cluster MS and binary stars we constructed
their respective radial density profiles by counting the stars distributed 
throughout the cluster region (see Fig.~\ref{fig:fig1}). Firstly, we split
the cluster area in small adjacent boxes of 
$0\fdg 10$$\times$$0\fdg 10$ that covered the entire analized field. Then, 
we counted the number of stars (MS and binary stars separately) inside them 
and  computed the mean densities as a function of the distance to the
 cluster's center ($d$) by averaging the star counts in every box
placed within annulus centered on the cluster with radii $d$ and  $d+\Delta d$).
This allowed us to estimate the uncertainties in star counts due to stellar
fluctuations within each annulus. We repeated this methodology using boxes of 
increasing size in steps of  $0\fdg 01$ per side, up to $0\fdg 20$$\times$$0\fdg 20$. The resulting radial density profiles shown in Fig.~\ref{fig:fig3} (left
panel) were obtained by averaging all the generated individual density radial profiles. We also built cumulative distributions as a function of $d$ (see Fig.~\ref{fig:fig3}, right panel) where the errors were calculated according to the Poisson statistics.
 For comparison purposes, although suffering from small number statistics, 
we also constructed the above curves for
cluster red giant stars ($G_0$ $<$ 13.8 mag, $(G_{BP} - G_{RP})_0$$>$1.0mag).

Figure~\ref{fig:fig3} reveals that cluster MS and binary stars are
distributed following an indistinguishable spatial pattern, from the cluster core
region out to the observed boundaries of its tidal tails  ($\sim$$1\fdg 4$
from the cluster' s center). As far as we are aware, binary stars 
populating cluster tidal tail regions out to $\sim$ 6 times the cluster
radius  ($\sim$ $0\fdg 22$ for UBC\,274) have not been observed in any 
Galactic open cluster yet.  Here we estimated the radius of the cluster
main body as the distance $r$ with the highest density contrast, calculated as
the ratio between the number of stars inside $r$ to the number of stars in an
annulus of interior radius $r$ and equal area. We refer the reader to some detailed studies of 
open clusters with well known tidal tails, namely: Hyades \citep{roseretal2019}, 
Praesepe \citep{rs2019}, Coma Berenices \citep{tangetal2019}, 
Blanco\,1\citep{zhangetal2020}, among others. Furthermore, until now, cluster
binary stars have been observed more centrally concentrated than single MS stars,
because of mass segregation caused by the dynamics internal evolution 
driven  by two-body relaxation 
\citep{reinoetal2018,gao2018,cohenetal2020}, even though open clusters
have also been subject of tidal effects due to the Milky Way gravitational potential
\citep{pm2018,piattietal2019b}. We therefore conclude that UBC\,274 has 
become the first open cluster with a numerous
binary star population  with a unique density profile.  The radial and cumulative
density profiles of red giant stars, although affected by small number 
statistics, also hint the removal of the signatures of the internal cluster dynamical evolution.

\begin{figure}
\includegraphics[width=\columnwidth]{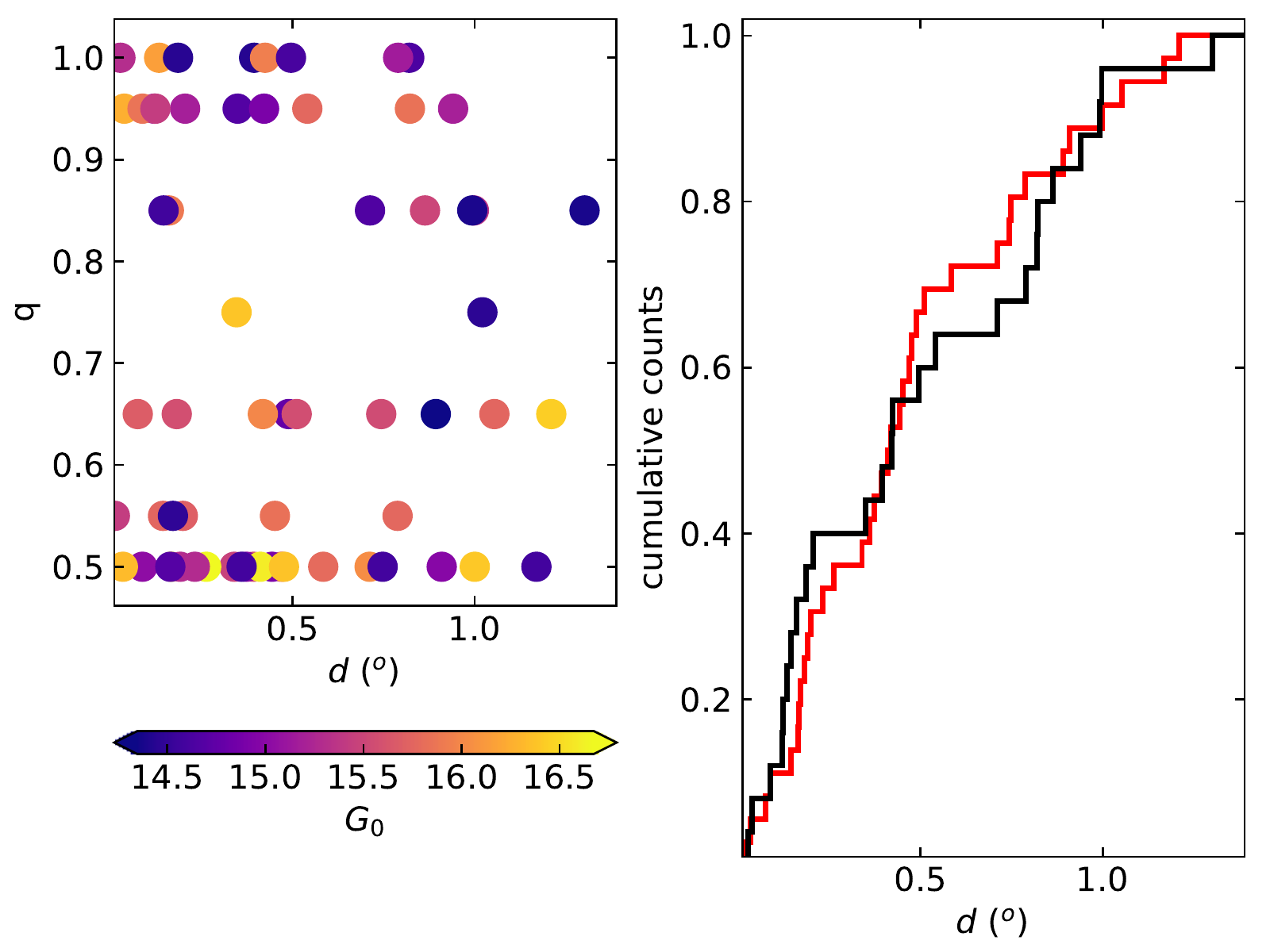}
\caption{Associated $q$ values to the cluster binary stars as a function of $d$,
 color-coded according to their intrinsic 
 $G_0$ magnitudes (left panel). The cumulative distribution for binary
stars with $q$ values smaller or larger than 0.75 are depicted with a red
and black line, respectively (right panel).}
\label{fig:fig4}
\end{figure}

We further investigated the spatial distribution of the ratio between the 
mass of the primary ($M_1$) and that of the secondary ($M_2$) binary star  
($M_2$/$M_1$=$q$, 0 $\le$ $q$ $\le$ 1). Following the precepts outlined by 
\citet{ht1998} and using the theoretical isochrone of log($t$ /yr)=9.45
\citep{betal12}, we computed the  $G_0$ magnitudes for binary stars
with $q$ values from 0.5 up to 1.0 in steps of 0.1 (see Fig.~\ref{fig:fig2},
right panel). We then interpolated the CMD positions of the binary stars
in the different generated binary star sequences in order to assign them the 
corresponding $q$ values.  Figure~\ref{fig:fig4} (left panel) shows the distribution
of $q$ values as a function of $d$ and  $G_0$ magnitudes. As can be seen, 
more massive binary stars (smaller  $G_0$ magnitudes) are found distributed 
along the entire $d$ range, irrespective of their $q$ values. This trend is
confirmed by the cumulative distribution functions built using binary stars
with $q$ values smaller or larger than 0.75 (see Fig~\ref{fig:fig4}, right
panel). Moreover, both binary star groups also seem to share similar
density profiles.

\begin{figure}
\includegraphics[width=\columnwidth]{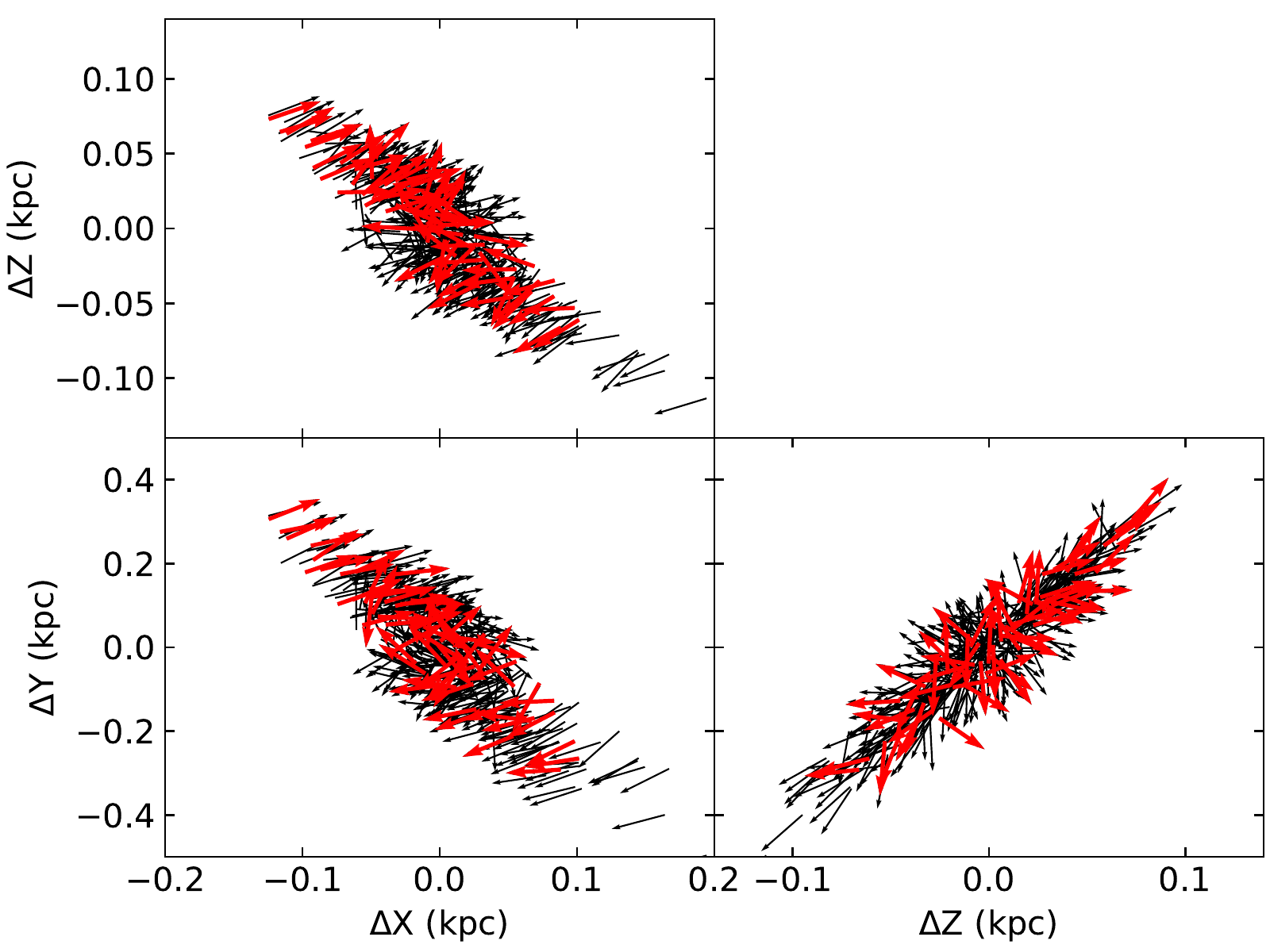}
\caption{Space velocity vectors with respect to the mean cluster motion 
projected on different Galactic planes  (relative XYZ coordinates) for cluster 
MS and binary stars, represented by black and red arrows, respectively. }
\label{fig:fig5}
\end{figure}

\begin{figure}
\includegraphics[width=\columnwidth]{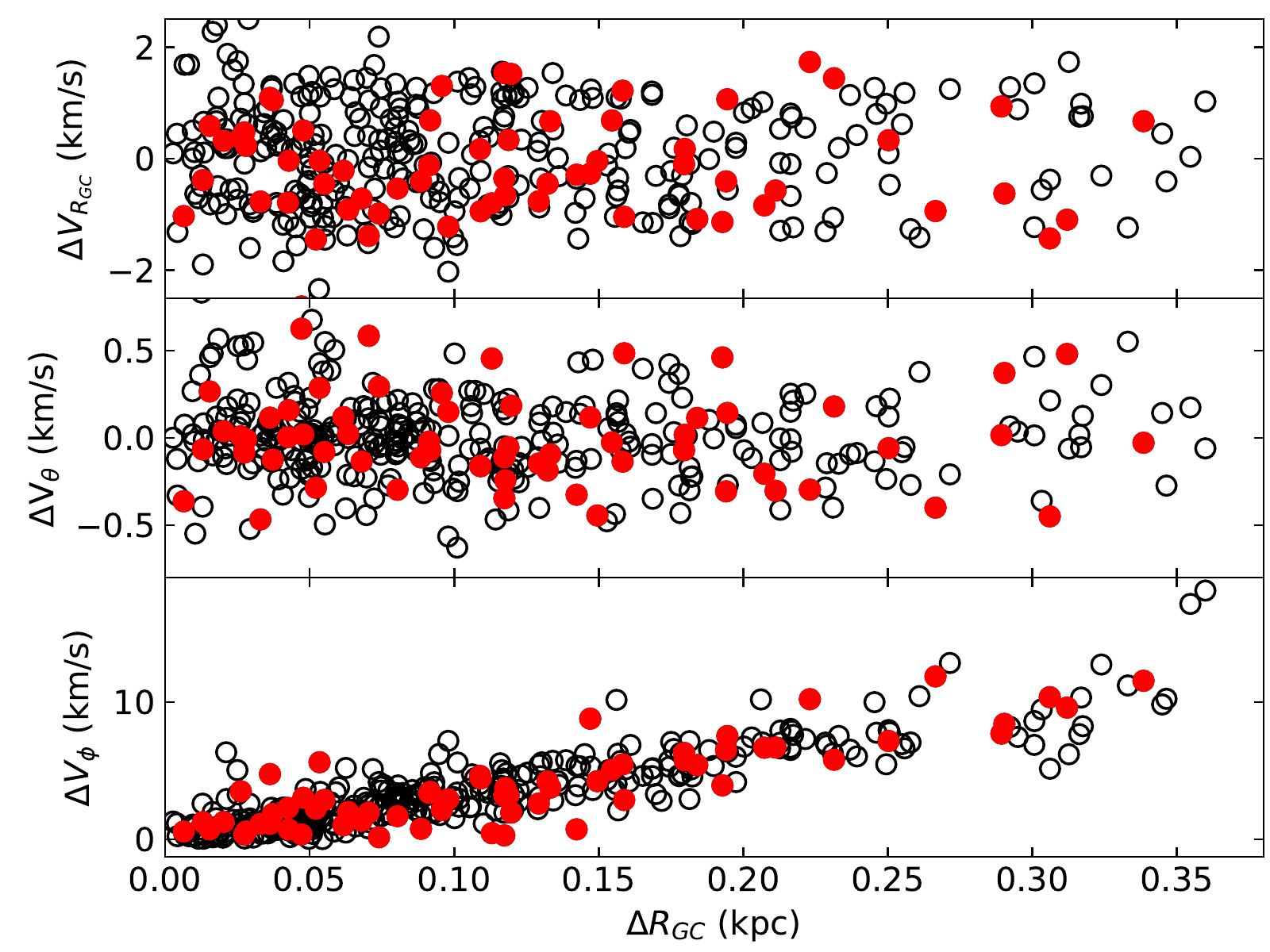}
\caption{Difference between individual space velocity components and those
of the mean cluster space velocity (spherical coordinate framework) as a 
function of  $R_{GC}$ (relative values). }
\label{fig:fig6}
\end{figure}

We finally analyzed the kinematics of cluster MS and binary stars from
the {\it Gaia} DR2 coordinates, proper motions, parallaxes and radial
velocities. Radial velocities are available for 13 cluster members.  For
cluster members without RV measurements, we assigned values
randomly generated in the range  [$<$RV$>$$-$$\sigma$(RV),$<$RV$>$$+$$\sigma$(RV)],
where $<$RV$>$ and $\sigma$RV are the cluster mean value and
dispersion obtained by \citet{castroginardetal2020}. We computed
Galactic coordinates $(X,Y,Z)$ and space velocities $(V_X,V_Y,V_Z)$
employing the \texttt{astropy}\footnote{https://www.astropy.org} package \citep{astropy2013,astropy2018}, which simply required the input of the astrometric
and kinematic data mentioned above.  Figure ~\ref{fig:fig5} illustrates the
space velocity vector field with respect to the mean cluster motion
projected on different Galactic planes (coordinates relative to the
cluster's center). The figure reveals that both cluster MS and binary stars
are moving remarkably fast with respect to the cluster's center all along the tidal
tails. In order to better understand that velocity pattern, we plotted
in Fig.~\ref{fig:fig6} the spherical components of the space velocities of the stars
with respect to the mean cluster velocity as a function of the relative
Galactocentric distances of them with respect to the that of the cluster's center.
From a kinematical point of view, Fig.~\ref{fig:fig6} provides further support 
that cluster MS and binary stars located at any position of the cluster main 
body or tidal tails have similar kinematical behaviors. Additionally it
shows that, while projection of the cluster motion on the plane of the
Galactic disk rotates like a solid body (bottom panel), the cluster is 
 being disrupted along the direction from the Galactic center (top panel) 
 and along the direction perpendicular to the Galactic plane (middle panel).

\section{Conclusions}

Motivated by the recent discovery of UBC\,274, an  $\sim$ 3 Gyr old
  open cluster located at $\sim$ 2 kpc from the Sun, which exhibits extended
tidal tails, we revisited the {\it Gaia} DR2 data set
for 365 cluster members.

From the cluster CMD corrected  for the differential reddening 
we recognized a numerous binary star population, distributed along the
whole magnitude dynamical range of the cluster MS and toward redder
colors. This binary star population spans $q$ values from 0.5 up to 1.0,
obtained by interpolation in a set of generated binary star sequences
from the theoretical isochrone computed by \citet{betal12} for the
cluster's age. We point out that the binary star strip neither is recognized 
from the observer cluster CMD nor from that corrected  for the the mean cluster
reddening.

We built stellar density radial profiles and cumulative distribution functions from 
the cluster's center out to $\sim$ 6 times  its radius for cluster MS and binary 
stars, separately.  The resulting radial profiles and cumulative distributions
show that binary stars reach the outermost regions of the observed tidal
tails and follow a similar spatial distribution pattern like cluster MS stars.
Moreover, in terms of spatial distribution, binary stars with $q$ values smaller or
larger than 0.75 do not show any difference either. The outcome that
a relatively large number of binary stars populates the cluster tidal tails could
become an observational evidence that brings support to the idea that a 
percentage of the field binary stars comes from disrupted star clusters 
\citep{goodwin2010}. 

As far as the kinematics of cluster members is considered, we computed
Galactic coordinates and space velocity components (cartesian and spherical 
frameworks) from the astrometric and kinematic information available. 
We found that stars located along the cluster's tidal tails, irrespective of 
being cluster  MS or binary stars,  are experiencing a relative fast motion 
with respect to the cluster's center. Particularly, the projected stellar motions 
on the  Galactic plane with respect to the cluster's center resembles that of
a rotating solid body, while those along the direction from the Galactic
center and perpendicular to the Galactic plane suggest that the
clusters is being disrupted. The resulting disrupting directions are in very
good agreement with realistic N-body simulations of the orbit of star
clusters with tidal tails \citep[e.g.,][]{montuorietal2007}. The similarity found
in the spatial distributions and kinematic patterns of cluster MS and binary
stars reveals that UBC\,274 is facing an intense process of
disruption that has apparently swept out any signature of internal dynamics
evolution driven by two-body relaxation.

\begin{acknowledgements}
 I thank the referee for the thorough reading of the manuscript and
timely suggestions to improve it. 
I also acknowledge support from the Ministerio de Ciencia, Tecnolog\'{\i}a e Innovaci\'on Productiva (MINCyT) through grant PICT-201-0030. 
\end{acknowledgements}

%\bibliographystyle{aa}
%\bibliography{paper} % if your bibtex file is called paper.bib

%\input{paper.bbl}

\end{document}